\documentclass[prd,floatfix,superscriptaddress,showpacs,showkeys,twocolumn]{revtex4-1}

\usepackage{amsmath}
\usepackage{epsfig}
\usepackage{graphics}
\usepackage{color}
\usepackage{graphicx}
\usepackage[font={footnotesize,it}]{caption}
\usepackage[colorlinks=false,
            linkcolor=red,
            urlcolor=red,
            citecolor=blue]{hyperref}

\begin{document}
\title{Effect of Lorentz Symmetry Breaking on the Deflection of Light in a Cosmic String Spacetime}
\author{Kimet Jusufi}
\email{kimet.jusufi@unite.edu.mk}
\affiliation{Physics Department, State University of Tetovo, Ilinden Street nn, 1200,
Tetovo, Macedonia}
\affiliation{Institute of Physics, Faculty of Natural Sciences and Mathematics, Ss. Cyril and Methodius University, Arhimedova 3, 1000 Skopje, Macedonia}
\author{\.{I}zzet Sakall\i{}}
\email{izzet.sakalli@emu.edu.tr}
\affiliation{Physics Department, Eastern Mediterranean University, Famagusta, Northern
Cyprus, Turkey}
\author{Ali \"{O}vg\"{u}n}
\email{ali.ovgun@pucv.cl}
\affiliation{Instituto de F\'{\i}sica, Pontificia Universidad Cat\'olica de Valpara\'{\i}%
so, Casilla 4950, Valpara\'{\i}so, Chile}
\affiliation{Physics Department, Eastern Mediterranean University, Famagusta, Northern
Cyprus, Turkey}
\date{\today }

\begin{abstract}
We investigate the Lorentz symmetry breaking effects (LSBE) on the
deflection of light by a rotating cosmic string spacetime in the weak limit
approximation. We first calculate the deflection angle by a static cosmic
string for a fixed spacelike 4-vector case (FSL) with the corresponding
effective-string optical metric using the Gauss-Bonnet theorem (GBT). Then, we focus on
a more general scenario, namely we calculate the deflection angle by a rotating
cosmic string applying the GBT to Randers effective-string metric. We
obtain a significant modification in the deflection angle because of the
LSBE parameter. We find a first and second order correction terms due to the global
effective topology which are proportional to the cosmic string and LSBE parameter, respectively. Finally, for a fixed time-like 4-vector (FTL) case, we show that the deflection angle is not affetced by LSBE parameter.
\end{abstract}

\pacs{95.30.Sf, 98.62.Sb, 04.40.Dg, 02.40.Hw}
\keywords{Light deflection; Gauss-Bonnet theorem; Cosmic String; Lorentz
symmetry breaking}
\maketitle

\section{Introduction}

Today, gravitational lensing phenomenon, which studies the effects of light
deflection on the appearance of cosmic objects is a useful theoretical tool
in observational cosmology \cite{sch}. Since Eddington's famous
observed eclipse \cite{edd}, the topic of gravitational lensing has gained
more attention. One of the remarkable information about the gravitational
field is that the deflection angle depends neither on the nature of the
matter nor on its physical state. Moreover, light deflection is an indirect
perception to compute the total matter density around a black hole \cite%
{glensing}. Besides, gravitational lensing can also shed light on the
possible existence and properties of topological defects that are formed
during the phase transitions in the early universe. The examples of those
topological defects are monopoles, cosmic strings and domain walls \cite%
{monopole1, monopole2}.

Recently, behaviour of a relativistic spin-0 particles that are subject to a
scalar potential under the effects of the Lorentz symmetry breaking \cite%
{bak1,bak2,bak3,bak4,bak5} in the cosmic string spacetime has been studied
by Bakke et al. \cite{bakke0,bakke1}. To tackle with the problem, they have
considered two possible scenarios of the anisotropy generated by a Lorentz
symmetry breaking effect (LSBE). They defined the LSBE term by using a
tensor $(\mathcal{K}_{F}){_{\mu \nu \alpha \beta }}$. This tensor governs
the Lorentz symmetry violation in the CPT [C: charge conjugation, P: parity,
and T: time reversal]-even gauge sector of the standard model extension \cite%
{CPTeven}. Using the scalar potential, which modifies the mass term in the
Klein- Gordon equation (KGE), it has been shown that the cosmic string
spacetime can be changed with the effects of the Lorentz symmetry violation
backgrounds.

An effective geometrical method for computing the gravitational lensing of a
considered black hole was introduced by Gibbons and Werner (GW) \cite%
{gibbons1,gibbons2}. GW's method calculates the asymptotic deflection angle
by employing the Gauss-Bonnet theorem (GBT). Moreover, the method of GW was
extended to the stationary metrics by Werner. Thus, he managed to get the
deflection angle for the Kerr black hole whose optical geometry is
Finslerian \cite{werner}. Today, there exists numerous papers in the
literature that use the GW's method. Along this line of thinking, for
example, the deflection angles of static cosmic strings and global monopoles
were studied by Jusufi \cite{kimet1,kimet2}. Furthermore, recently the
deflection angle for the infra-red region by using the Gaussian curvature of
the optical metric of Rindler modified Schwarzschild back hole has been
investigated by Sakalli and Ovgun \cite{sakalli} in which the role of the
Rindler acceleration on the gravitational lensing is neatly shown.
Meanwhile, it is worth noting that by applying a complex coordinate
transformation, Newman and Janis \cite{newman} established a relationship
between the non-rotating and rotating spacetimes of general relativity.

In this paper, we use the cosmic string spacetime with the LSBE \cite%
{bakke0,bakke1} to analyze the light deflection predicted by Einstein's
General Theory of Relativity. To this end, we follow the method of GW \cite%
{gibbons1,gibbons2}. Thus, by integrating the Gaussian curvature of the
optical metric outwards from the light ray, we plan to reveal how the LSBE\
plays role on the cosmic string spacetime and modifies the deflection angle.

The paper is organized as follows. In Sec. II, we briefly review the FSL 4-vector case of the cosmic string metric of \cite%
{bakke0,bakke1}. The second part of this section is devoted to the derivation of effective
Gaussian curvature within\ the GBT and its corresponding deflection angle.
In Sec. III, we study the deflection angle of the rotating cosmic string
metric with LSBE. In particular we consider the effective Gaussian optical curvature and deflection angle. In Sec. IV, we briefly review and calculate the deflection angle for a fixed FTL 4-vector case of a static cosmic string with the LSBE. In Sec. VII, we extend our results and investigate the deflection angle for a rotating cosmic string for a FTL 4-vector case.  In Sec. V, we  consider the geodesics equations to recover the deflection angles. Finally, we draw our conclusions in Sec. VI. Throughout this paper, we  shall use natural units, i.e. $G=c=\hbar=1$.

\section{Effective cosmic string metric for a FSL 4-vector case}
\subsection{A static cosmic string for a FSL 4-vector case}
The Lagrange density for the non-birefringent modified Maxwell theory
coupled to gravity \cite{bakke0,Kant1,Kant2} is given by 
\begin{equation}
\mathcal{L}_{mod}=-\sqrt{-g}\left( \frac{1}{4}F_{\mu \nu }F_{\rho \sigma
}g^{\mu \rho }g^{\nu \sigma }+\frac{1}{4}\mathcal{K}^{\mu \nu \rho \lambda
}F_{\mu \nu }F_{\rho \lambda }\right) ,  \label{1}
\end{equation}

where $\mathcal{K}^{\mu \nu \rho \lambda }$ is a Lorentz symmetry violating
tensor, which guarantees the CPT symmetry. $\mathcal{K}^{\mu \nu \rho
\lambda }$ shares the same features of the Riemann tensor, plus some
additional double-traceless conditions: 
%
%
%
%
\begin{equation}
\mathcal{K}_{\mu \nu \rho \lambda }=\mathcal{K}_{[\mu \nu ][\rho \lambda ]},%
\enspace\mathcal{K}_{\mu \nu \rho \lambda }=\mathcal{K}_{\rho \lambda \mu
\nu },\enspace\mathcal{K}_{\>\enspace\mu \nu }^{\mu \nu }=0.  \label{2}
\end{equation}%
By using the following effective metric tensor 
\begin{equation}
g^{\text{eff}}_{\mu \rho }(x)=g_{\mu \rho }(x)+\epsilon \xi _{\mu }\xi _{\rho },
\label{3}
\end{equation}%
in which the parameter $\epsilon $ is governed by $\epsilon =\kappa
/(1+\kappa \,\xi_{\rho} \,\xi^{\rho} /2)$ with $0\leq \kappa <2$, one can see (as being stated in \cite%
{bakke0}) that $g^{\mu \nu }(x)$ background attributes an anisotropy, which
means that the propagation of light must be modified by the background. Let
us first consider a normalized parameter four-vector $\xi _{a}$ as a
space-like 4-vector: 
\begin{equation}
\xi _{a}=\left( 0,0,1,0\right) .  \label{4}
\end{equation}%
Under the Lorentz symmetry breaking, a topological defect in curved
spacetime can be expressed by the effective metric tensor of the cosmic
string \cite{Xan}, whose line-element in cylindrical coordinates is given by the effective metric
\cite{bakke0} 
\begin{equation}\label{5}
\mathrm{d}s^{2}=-\mathrm{d}t^{2}+\mathrm{d}\rho ^{2}+\eta ^{2}\rho
^{2}\left( 1+\epsilon \right) \mathrm{d}\varphi ^{2}+\mathrm{d}z^{2}.
\end{equation}
In the above equation $\eta $ is the parameter of the cosmic string.
Moreover, $\eta $ is expressed by $\eta =1-4\mu $, where $\mu $ is the
linear mass density of the cosmic string. We can easily write the above
metric in spherical coordinates. To do so, let us introduce the following
coordinates transformations $z=r\cos \theta $ and $\rho =r\sin \theta $.
Thus, metric \eqref{5} becomes 
\begin{equation}
\mathrm{d}s^{2}=-\mathrm{d}t^{2}+\mathrm{d}r^{2}+r^{2}\mathrm{d}\theta ^{2}+\eta
^{2}r^{2}\left( 1+\epsilon \right) \sin ^{2}{\theta }\mathrm{d}\varphi ^{2},
\label{6}
\end{equation}%
then we can find the corresponding optical metric form of metric \eqref{6}, if we first
project the metric into equilateral plane with $\theta =\pi /2$ and
immediately after we consider the null case: $\mathrm{d}s^{2}=0$. Therefore, one gets
the optical metric of the line-element \eqref{6} as follows 
\begin{equation}
\mathrm{d}t^{2}=\mathrm{d}r^{2}+\eta ^{2}r^{2}\left( 1+\epsilon \right) 
\mathrm{d}\varphi ^{2}.  \label{7}
\end{equation}%
We now introduce a new coordinate $r^{\star }$, thereby a new function $%
f(r^{\star })$: 
\begin{equation}
\mathrm{d}r^{\star }=\mathrm{d}r,\,\,\,f(r^{\star })=\eta r\sqrt{1+\epsilon }%
.  \label{8}
\end{equation}%
Therefore, the optical metric \eqref{7} becomes \cite{gibbons1}
\begin{equation}
\mathrm{d}t^{2}=\tilde{g}_{ab}\,\mathrm{d}x^{a}\mathrm{d}x^{b}=\mathrm{d}{%
r^{\star }}^{2}+f^{2}(r^{\star })\mathrm{d}\varphi ^{2}.  \label{9}
\end{equation}%
To derive the corresponding effective Gaussian optical curvature $K$ of the metric
\eqref{9}, we follow \cite{gibbons1}: 
\begin{eqnarray}
K &=&-\frac{1}{f(r^{\star })}\frac{\mathrm{d}^{2}f(r^{\star })}{\mathrm{d}{%
r^{\star }}^{2}}, \notag \\
&=&-\frac{1}{f(r^{\star })}\left[ \frac{\mathrm{d}r}{\mathrm{d}r^{\star }}%
\frac{\mathrm{d}}{\mathrm{d}r}\left( \frac{\mathrm{d}r}{\mathrm{d}r^{\star }}%
\right) \frac{\mathrm{d}f}{\mathrm{d}r}+\left( \frac{\mathrm{d}r}{\mathrm{d}%
r^{\star }}\right) ^{2}\frac{\mathrm{d}^{2}f}{\mathrm{d}r^{2}}\right] .
\label{10}
\end{eqnarray}%
Since $f(r^{\star })$ is linear in $r$,  one can check by using Eq. \eqref{10} that in fact the effective Gaussian optical vanishes, i.e. $K=0$. This result will provide us some convenience during the
computation of the light deflection in the following sections.

\subsection{Effective Gaussian Curvature and Deflection angle}

The GBT for the non-singular region $D_{R}$ in $M$, with boundary $\partial
D_{R}=\gamma _{\tilde{g}}\cup C_{R}$ can be stated as follows \cite{gibbons1}
\begin{equation}
\iint\limits_{\mathcal{D}_{R}}K\,\mathrm{d}S+\oint\limits_{\partial \mathcal{%
D}_{R}}\kappa \,\mathrm{d}t+\sum_{i}\theta _{i}=2\pi \chi (\mathcal{D}_{R}).
\label{11}
\end{equation}

Note that $\kappa $ is the geodesic curvature, $K$ gives the Gaussian optical curvature, $\theta_{i}$ gives the corresponding exterior angle at the $i^{th}$ vertex, and $%
\chi (\mathcal{D}_{R})$ is the Euler characteristic number. The geodesic curvature can be computed as $\kappa =\tilde{g}\,(\nabla _{\dot{%
\gamma}}\dot{\gamma},\ddot{\gamma})$, in which the unit speed condition holds $\tilde{g}(\dot{\gamma},\dot{%
\gamma})=1$ and $\ddot{\gamma}$ is the unit acceleration vector.

While $R\rightarrow \infty $, both jump angles ($\theta _{\mathit{O}}$ , $%
\theta _{\mathit{S}}$) become $\pi /2,$ and hence $\theta _{\mathit{O}%
}+\theta _{\mathit{S}}\rightarrow \pi $; in which the subscripts $\mathit{S}$
and $\mathit{O}$ correspond to the source and observer, respectively (see
for example \cite{gibbons1}). Furthermore, since $\gamma _{\tilde{g}}$ is a
geodesic, then it follows $\kappa (\gamma _{\tilde{g}})=0$. Let us find now
the geodesic curvature, which can be calculated as $\kappa (C_{R})=|\nabla _{%
\dot{C}_{R}}\dot{C}_{R}|$ in which one can choose $C_{R}:=r(\varphi
)=R=\text{const}.$ The radial component of the geodesic curvature can be calculated
as 
\begin{equation}
\left( \nabla _{\dot{C}_{R}}\dot{C}_{R}\right) ^{r}=\dot{C}_{R}^{\varphi
}\,\left( \partial _{\varphi }\dot{C}_{R}^{r}\right) +\tilde{\Gamma} _{\varphi
\varphi }^{r}\left( \dot{C}_{R}^{\varphi }\right) ^{2}.  \label{12}
\end{equation}

The first term vanishes, while the second term gives nonzero contribution. To see this, we  should recall the nonzero component $\tilde{\Gamma} _{\varphi \varphi }^{r}=-f(r^{\star })f^{\prime }(r^{\star })
$ and the unit speed condition  $\tilde{%
g}_{\varphi \varphi }\, \dot{C}_{R}^{\varphi }\dot{C}_{R}^{\varphi }=1$, where $f(r^{\star })$ is given by Eq. \eqref{8}. Using these relations and the optical metric \eqref{9}, it follows immediately that
\begin{equation}
\mathrm{d}t=\eta R\sqrt{1+\epsilon }\,\mathrm{d}\varphi .  \label{13}
\end{equation}

Thus, for very large $R,$ the geodesic curvature reads  
\begin{eqnarray}
\lim_{R\rightarrow \infty }\kappa (C_{R}) &=&\lim_{R\rightarrow \infty
}\left\vert \nabla _{\dot{C}_{R}}\dot{C}_{R}\right\vert ,  \notag \\
&=&\lim_{R\rightarrow \infty }\left( \frac{\eta ^{2}(1+\epsilon )}{\eta
^{2}R^{2}(1+\epsilon )}\right) ^{1/2},  \notag \\
&\rightarrow &\frac{1}{R},  \label{14}
\end{eqnarray}%
which suggests that

\begin{equation}
\kappa (C_{R})\mathrm{d}t=\eta \sqrt{1+\epsilon }\mathrm{d}\,\varphi .
\label{15n}
\end{equation}

Reconsidering Eq. \eqref{11} and recalling that the Euler characteristic is
characterized by $\chi (\mathcal{D}_{R})=1$, we find 
\begin{equation}
\iint\limits_{\mathcal{D}_{R}}K\,\mathrm{d}S+\oint\limits_{C_{R}}\kappa \,%
\mathrm{d}t\overset{{R\rightarrow \infty }}{=}\iint\limits_{\mathcal{D}%
_{\infty }}K\,\mathrm{d}S  \notag
\end{equation}

\begin{equation}
+\eta \sqrt{1+\epsilon }\int\limits_{0}^{\pi +\hat{\alpha}}\mathrm{d}\varphi
=\pi,  \label{16}
\end{equation}

in which the domain $\mathcal{D}_{\infty }$ connotates an infinite domain
bounded by the light ray $\gamma _{\tilde{g}}.$ Thus, the asymptotic
deflection angle $\hat{\alpha}$\ can be found as 
\begin{equation}
\hat{\alpha}=\frac{\pi }{\eta \sqrt{1+\epsilon }}-\pi .  \label{17}
\end{equation}

Using Taylor series in $\eta$ and $\epsilon$, we can approximate the result for the deflection angle as
\begin{equation}
\hat{\alpha}\simeq 4\mu \pi -\frac{\epsilon \pi }{2}-2\pi \mu \epsilon +%
\mathcal{O}(\mu ^{2},\eta ^{2}).  \label{18}
\end{equation}

The first term is just the deflection angle by a static cosmic string. Interestingly, due to the Lorentz symmetry breaking by the parameter $\epsilon$, we find that the deflection angle decreases.

\section{Rotating cosmic strings with FSL 4-vector}
\subsection{Effective String-Randers optical metric}
Let us introduce a rotating cosmic string by using the following coordinate
transformations \cite{mazur,bezerra}
\begin{equation}
\mathrm{d}t\rightarrow \mathrm{d}t+a\,\mathrm{d}\varphi ,  \label{19s}
\end{equation}%
into the metric \eqref{5}, we find
\begin{equation}\label{20}
\mathrm{d}s^{2}=-\left(\mathrm{d}t+a\,\mathrm{d}\varphi\right)^{2}+\mathrm{d}\rho ^{2}+\eta ^{2}\rho
^{2}\left( 1+\epsilon \right) \mathrm{d}\varphi ^{2}+\mathrm{d}z^{2}.
\end{equation}

Let us now introduce the tetrads ${e^a}_{\mu}(x)$, which satisfies the relation $g_{\mu\nu}(x)={e^a}_{\mu}(x){e^b}_{\mu}(x) \eta_{ab}$, in which $\eta_{ab}$ is the Minkowski tensor. In particular we choose the tetrads as follows
\begin{equation}
{e^a}_{\mu}(x)=\begin{pmatrix}
 1 & 0 &  0& 0\\ 
 0&  1&  0& 0\\ 
a &  0&  \eta \rho \sqrt{1+\epsilon}& 0\\ 
 0& 0 &  0& 1
\end{pmatrix}.
\end{equation}

Next, by writing the four vector $\xi_{\mu}(x)$, in terms of tetrads as $\xi_{\mu}(x)={e^a}_{\mu}(x) \xi_{a}$, and choosing $\xi_{a}=(\zeta,\sigma,\gamma,\delta)$, one can show that $\xi_{\mu}(x) \xi^{\mu}(x) =-\zeta^2+\sigma^2+\gamma^2+\delta^2=\text{const}$ holds for the rotation case. In particular, the choice of $\xi_{a}=(0,0,1,0)$ results in $\xi_{\mu}(x) \xi^{\mu}(x) =1=\text{const}$. Metric \eqref{20} can be expressed in spherical coordinates as follows
\begin{equation}
\mathrm{d}s^{2}=-(\mathrm{d}t+a\mathrm{d}\varphi )^{2}+\mathrm{d}r^{2}+r^{2}%
\mathrm{d}\theta ^{2}+\eta ^{2}r^{2}\alpha \sin ^{2}{\theta }\mathrm{d}%
\varphi ^{2},  \label{22}
\end{equation}%
where we have introduced $\alpha =1+\epsilon $. The stationary metric can be recasted to give a Finslerian optical metric of Randers type with the Hessian given as \cite{werner} 
\begin{equation}
g_{ij}(x,X)=\frac{1}{2}\frac{\partial ^{2}F^{2}(x,X)}{\partial X^{i}\partial
X^{j}}.  \label{23}
\end{equation}

Furthermore, by homogeneity we have $F^{2}(x,X)=g_{ij}(x,X)X^{i}X^{j}$, thus the Randers metric can also be written in the following from \cite{gibbons2}
\begin{equation}
F(x,X)=\sqrt{a_{ij}(x)X^{i}X^{j}}+b_{i}(x)X^{i},  \label{24}
\end{equation}%
where $a_{ij}$ and $b_{i}$ must satisfy the condition $a^{ij}b_{i}b_{j}<1$.

One can easily find the corresponding Randers optical metric for our stationary effective cosmic string spacetime by writing the stationary spacetime \eqref{22} as \cite{gibbons2} 
\begin{equation}\label{25}
\mathrm{d}s^{2}=V^{2}\left[ -\left( \mathrm{d}t-b_{i}\mathrm{d}x^{i}\right)
^{2}+a_{ij}\mathrm{d}x^{i}\mathrm{d}x^{j}\right],  
\end{equation}%
where
\begin{eqnarray}
a_{ij}(x)\mathrm{d}x^{i}\mathrm{d}x^{j} &=&\mathrm{d}r^{2}+r^{2}\mathrm{d}%
\theta ^{2}+\eta ^{2}r^{2}\alpha \sin ^{2}\theta \mathrm{d}\varphi ^{2}, 
 \\
b_{i}(x)\mathrm{d}x^{i} &=&-a\mathrm{d}\varphi ^{2}.  
\end{eqnarray}%

In this paper we shall consider a planar light ray by setting $\theta =\pi /2$, then making use of Eqs. \eqref{24} and \eqref{25}, we end up with the effective Randers-string metric: 
\begin{equation}
F\left( r,\varphi ,\frac{\mathrm{d}r}{\mathrm{d}t},\frac{\mathrm{d}\varphi }{%
\mathrm{d}t}\right) =\sqrt{\left( \frac{\mathrm{d}r}{\mathrm{d}t}\right)
^{2}+\eta ^{2}r^{2}\alpha \left( \frac{\mathrm{d}\varphi }{\mathrm{d}t}%
\right) ^{2}}-a\frac{\mathrm{d}\varphi }{\mathrm{d}t}.  \label{28}
\end{equation}

Hence the Randers metric for null geodesics $\mathrm{d}s^{2}=0$, gives $\mathrm{d}t=F(x,\mathrm{d}x)$.  On the other hand Fermat's principle suggests that light rays  $\gamma $ are selected by the following condition 
\begin{equation}
0=\delta \,\int\limits_{\gamma }\mathrm{d}t=\delta \,\int\limits_{\gamma
_{F}}F(x,\dot{x})\mathrm{d}t,  \label{29}
\end{equation}%
where it is important to note that these spatial light rays $\gamma $ are also geodesics $\gamma _{F}$ of the Randers metric $F$. This is crucial point since we can  apply the so-called Naz{\i }m's method \cite{nazim} to construct a
Riemannian manifold $(\mathcal{M},\bar{g})$, osculating the Randers manifold 
$(\mathcal{M},F)$. To do so, we choose a smooth and non-zero vector field $\bar{X}$ over $%
\mathcal{M}$  (except at single vertex points) with $\bar{%
X}(\gamma _{F})=\dot{x}$. The Hessian \eqref{23} then reads 
\begin{equation}
\bar{g}_{ij}(x)=g_{ij}(x,\bar{X}(x)).  
\end{equation}

It is quite remarkable fact that the geodesic $\gamma _{F}$ of $(\mathcal{M},F)$
is also a geodesic $\gamma _{\bar{g}}$ of $(\mathcal{M},\bar{g})$ i.e., $%
\gamma _{F}=\gamma _{\bar{g}}$ (see \cite{werner} for details). Hence we shall use the cosmic string effective optical metric and construct the corresponding
osculating Riemannian manifold $(\mathcal{M},\bar{g})$. This is important since allows us to calculate the deflection angle of the planar light ray. We choose 
the line $r(\varphi )=b/\sin \varphi $, where $b$ is known as the impact parameter and gives the minimal radial distance of the light ray from the cosmic string lying along the $z$ axis.  We make  the following choose for the leading terms of the vector field $\bar{X}=(\bar{X}^{r},\bar{X}^{\varphi })(r,\varphi )$ 
(see for detales \cite{werner}) 
\begin{equation}
\bar{X}^{r}=-\cos \varphi +\mathcal{O}(a),\hspace{1cm}\bar{X}^{\varphi }=%
\frac{\sin ^{2}\varphi }{b}+\mathcal{O}(a).  \label{31}
\end{equation}

In the next section we shall use these relationsto calculate the Gaussian curvature and apply the GBT theorem to the osculating Riemannian optical metric.

\subsection{Deflection angle}
%

In this section we shall apply the GBT to the osculating optical geometries. In particular one can apply the GBT to a domain $(\mathcal{D}_{R},\bar{g})$ within the
region $\mathcal{D}_{R}$ possessing boundary curve $\partial \mathcal{D}%
_{R}=\gamma _{\bar{g}}\cup C_{R}$ \cite{werner} 
\begin{equation}
\iint\limits_{\mathcal{D}_{R}}K\,\mathrm{d}S+\oint\limits_{\partial \mathcal{%
D}_{R}}\kappa \,\mathrm{d}t+\sum_{i}\theta _{i}=2\pi \chi (\mathcal{D}_{R}).
\label{32}
\end{equation}%

Note that in a similar way we define the geodesic curvature as $\kappa =|\nabla _{\dot{\gamma}}\dot{%
\gamma}|$ (with respect to $\bar{g}$).
Hereinbefore, as $R\rightarrow \infty $ the both jump angles
tends to $\pi /2$, hence in a analogues way as in the last section we have $\theta _{O}+\theta _{S}\rightarrow \pi $.
Thus, Eq. \eqref{32} recasts in 
\begin{equation}
\iint\limits_{\mathcal{D}_{R}}K\,\mathrm{d}S+\oint\limits_{\partial \mathcal{%
D}_{R}}\kappa \,\mathrm{d}t=2\pi \chi (\mathcal{D}_{R})-(\theta _{O}+\theta
_{S})=\pi .  \label{33}
\end{equation}

Since the geodesic curvature in the case of geodesics $\gamma _{\bar{g}}$
vanishes i.e. $\kappa (\gamma _{\bar{g}})=0$, we shall now focus on
calculating $\kappa (C_{R})\mathrm{d}t$ where $\kappa (C_{R})=|\nabla _{\dot{%
C}_{R}}\dot{C}_{R}|$. For very large but constant $R$ given by $%
C_{R}:=r(\varphi )=R=\text{const}$, if we use the unit speed condition i.e. $\bar{g}_{\varphi \varphi
}\,\dot{C}_{R}^{\varphi }\dot{C}_{R}^{\varphi }=1$, and the nonzero Christoffel
symbol $\bar{\Gamma}_{\varphi \varphi }^{r}$, the geodesic
curvature is found to be $\kappa (C_{R})\rightarrow R^{-1}$. The effective cosmic string-optical metric \eqref{28} then gives 
\begin{equation}
\mathrm{d}t=\left( \sqrt{\eta ^{2}R^{2}\alpha }-a\right) \mathrm{d}\,\varphi
.  \label{34}
\end{equation}

Hence if we combine these results it follows
\begin{equation}
\lim_{R\rightarrow \infty }\kappa (C_{R})\mathrm{d}t=\lim_{R\rightarrow
\infty }\left( \sqrt{\eta ^{2}\alpha }-\frac{a}{R}\right) =\eta \sqrt{\alpha 
}\,\mathrm{d}\,\varphi .  \label{35}
\end{equation}

We clearly see that our effective optical metric is not asymptotically
Euclidean i.e. $\kappa (C_{R})\mathrm{d}t/\mathrm{d}\varphi =\eta \sqrt{%
\alpha }\neq 1$, due to the fact that our spacetime metric \eqref{22} is globally conical. Clearly if we set $\eta \sqrt{\alpha }\rightarrow 1$, we find the asymptotically
Euclidean case, i.e. $\kappa (C_{R})\mathrm{d}t/\mathrm{d}\varphi =1$. To find the deflection angle  we first approximate the boundary curve of $\mathcal{D}%
_{\infty }$ by a notional undeflected ray, that is, the line $r(\varphi
)=b/\sin \varphi $, then GBT reduces to 
\begin{equation}
\hat{\alpha}\simeq \pi \left( \frac{1}{\eta \sqrt{\alpha }}-1\right) -\frac{1%
}{\eta \sqrt{\alpha }}\int\limits_{0}^{\pi }\int\limits_{\frac{b}{\sin
\varphi }}^{\infty }K\,\sqrt{\det \bar{g}}\,\mathrm{d}r\,\mathrm{d}\varphi .
\label{36}
\end{equation}

We can make use of the Eqs. \eqref{23}, \eqref{33}, and \eqref{35}, to find the following relations
\begin{align}\label{39}
\bar{g}_{rr}&=1-\frac{ \sin^{6}\varphi \alpha \eta^2 r^2 a }{b^3 \left(\cos^{2}\varphi+\frac{r^2 \eta^{2} \alpha \sin^{4}\varphi }{b^2}\right)^{3/2}}+\mathcal{O}(a^2),\\
\bar{g}_{\varphi \varphi}&=r^2 \eta^2 \alpha-\frac{a\sin^2 \varphi \alpha r^2 \left(2 r^2 \eta^2 \alpha \sin^4 \varphi +3 b^2 \cos^2 \varphi \right)\eta^2 }{b^3 \left(\cos^{2}\varphi+\frac{r^2 \eta^{2} \alpha \sin^{4}\varphi }{b^2}\right)^{3/2}}\\
\bar{g}_{r\varphi}&=\frac{a \cos^{3}\varphi}{b^3 \left(\cos^{2}\varphi+\frac{r^2 \eta^{2} \alpha \sin^{4}\varphi }{b^2}\right)^{3/2}}+\mathcal{O}(a^2),
\end{align}
neglecting higher order terms of the angular momentum parameter $a$.  Then the determinant of this metric can be written as
\begin{equation}
\det \bar{g}=r^2 \eta^2 \alpha- \frac{3 a\sin^2 \varphi \alpha \eta^2 r^2 (\sin^4 \varphi \alpha \eta^2 r^2 +\cos^2 \varphi b^2)}{b^3 \left(\cos^{2}\varphi+\frac{r^2 \eta^{2} \alpha \sin^{4}\varphi }{b^2}\right)^{3/2}}
\end{equation}
For the Christoffel symbols we find,
\begin{eqnarray}\notag
\bar{\Gamma}^{\varphi}_{rr}&=&-\frac{3 a \sin^4\varphi \left[2(-\sin\varphi r+b)\cos^2\varphi -r \sin^3 \varphi \right] \cos\varphi}{2 r b^3 \left(\cos^{2}\varphi+\frac{r^2 \eta^{2} \alpha \sin^{4}\varphi }{b^2}\right)^{5/2}},\\
\bar{\Gamma}^{\varphi}_{r\varphi}&=&\frac{1}{r}+\frac{a \alpha \eta^2 r \sin^6 \varphi \left[2 \sin^4 \varphi \alpha \eta^2 r^2+5 b^2 \cos^2 \varphi\right]}{2 \left(\cos^{2}\varphi+\frac{r^2 \eta^{2} \alpha \sin^{4}\varphi }{b^2}\right)^{5/2} b^5}.
\end{eqnarray}
The Gaussian curvature is
\begin{eqnarray}
K&=&\frac{\bar{R}_{r\varphi r\varphi}}{\det \bar{g}}\\\notag
&=&\frac{1}{\sqrt{\det \bar{g}}}\left[\frac{\partial}{\partial \varphi}\left(\frac{\sqrt{\det \bar{g}}}{\bar{g}_{rr}}\,\bar{\Gamma}^{\varphi}_{rr}\right)-\frac{\partial}{\partial r}\left(\frac{\sqrt{\det \bar{g}}}{\bar{g}_{rr}}\,\bar{\Gamma}^{\varphi}_{r\varphi}\right)\right],
\end{eqnarray}
so using the Christoffel symbols and the metric components, we obtain
\begin{equation}
K=-\frac{12 a }{r }f(r,\varphi,\eta,\alpha),\label{26}
\end{equation}
with
\begin{widetext}
\begin{eqnarray}\notag
f(r,\varphi,\eta,\alpha)&=& \frac{\sin^3 \varphi}{\left(\cos^{2}\varphi+\frac{r^2 \eta^{2} \alpha \sin^{4}\varphi }{b^2}\right)^{7/2}b^7}\Big[-\frac{\sin^{11}\varphi  \alpha^3 \eta^6 r^5}{24}+\frac{b^2 r^3 \eta^2 \alpha \sin^9 \varphi}{8}+\frac{r^3 \eta^2 \alpha \cos^{2}\varphi b^2 (\eta^2 \alpha+27) \sin^7 \varphi}{24} \\
&-&\frac{3 b^3 r^2 \eta^2 \alpha \cos^2 \varphi \sin^6 \varphi }{4}+\Big(\frac{5 b^2 r^3 \eta^2 \alpha \cos^4 \varphi}{4}-\frac{\cos^2 \varphi b^4 r}{2}\Big)\sin^5 \varphi- \frac{3 r^2 b^3 \alpha \eta^2 \cos^4 \varphi \sin^4 \varphi }{2}\\
&+& \frac{17 r (\eta^2 \alpha -\frac{33}{17}) \cos^4 \varphi b^4 \sin^3 \varphi}{24}+\frac{\cos^4 \varphi \sin^2 \varphi b^5}{2} -\frac{5 b^4 r \sin \varphi \cos^6 \varphi}{4}+b^5 \cos^6 \varphi \Big].  
\end{eqnarray}
\end{widetext}

The deflection angle \eqref{36} reduces to
\begin{eqnarray}\notag
\hat{\alpha}&\simeq & 4\pi \mu-\frac{\epsilon \pi}{2}-2 \mu \epsilon \pi -\frac{1}{\eta \sqrt{\alpha}}\\
&\times &\int\limits_{0}^{\pi}\int\limits_{\frac{b}{\sin \varphi}}^{\infty}\left(-\frac{12a}{r}f(r,\varphi,\eta, \alpha)\right)\sqrt{\det \bar{g}}\,\mathrm{d}r\,\mathrm{d}\varphi,
\end{eqnarray}
in which $b$ is the impact parameter. After we integrate with respect to the radial coordinate and considering a Taylor expansion around $\mu$ and $\epsilon$, we find a non-zero contribution for a retrograde light ray
\begin{eqnarray}\notag
&&\int\limits_{0}^{\pi}\int\limits_{\frac{b}{\sin \varphi}}^{\infty} \frac{12  a}{r}f(r,\varphi,\eta,\alpha)\sqrt{\det \bar{g}}\,\mathrm{d}r\,\mathrm{d}\varphi \\
&=&\frac{3 a \pi \mu}{2 b}-\frac{3 a \pi \epsilon}{16 b}-\frac{3 a \pi \epsilon \mu }{4 b}+\mathcal{O}(\mu^2,\epsilon^2).
\end{eqnarray}

But zero contribution for the prograde light ray
\begin{eqnarray}
&&\int\limits_{0}^{\pi}\int\limits_{\frac{b}{\sin \varphi}}^{\infty} \frac{12  a}{r}f(r,\varphi,\eta,\alpha)\sqrt{\det \bar{g}}\,\mathrm{d}r\,\mathrm{d}\varphi=0.
\end{eqnarray}

Finally the deflection angle for the retrograde case gives
\begin{equation}\label{49}
\hat{\alpha}_{ret}\simeq 4\pi \mu-\frac{\epsilon \pi}{2}-2 \pi \mu \epsilon+\frac{3 a \pi \mu}{2 b}-\frac{3 a \pi \epsilon}{16 b}, 
\end{equation}
and similary for the  prograde case
\begin{equation}
\hat{\alpha}_{prog}\simeq 4\pi \mu-\frac{\epsilon \pi}{2}-2 \pi \mu \epsilon.
\end{equation}

It is interesting to note that since the rotating cosmic string parameter $a$ is proportional to the angular momentum, i.e. $a=4J$, which contains the mass per unit length, $\mu$, by definition of the angular momentum. As a consequence, the last two terms in Eq. \eqref{49} can be considered as second order terms, more precisely, the term $a \mu$ can be viewed as $\mu^2$, while the last term can also be viewed as $\mu^2$, is we assume $\mu$ and $\epsilon$, say, to be of the same order of magnitude. In this way, if we neglect these terms, we end up with the following result
\begin{equation}\label{51}
\hat{\alpha}\simeq 4\pi \mu-\frac{\epsilon \pi}{2}-2 \pi \mu \epsilon.
\end{equation}

\section{Effective cosmic string metric for a FTL 4-vector case with LSBE}
\subsection{Static cosmic strings}
In this section we shall consider a normalized parameter four-vector $\xi_a$ as a time-like
4-vector given by
\begin{equation}
\xi_a=\left(1,0,0,0\right).
\end{equation}

In this case, under the LSB, a cosmic string metric can be expressed by the effective metric \cite{Xan}, whose line-element in cylindrical coordinates is given by \cite{bak1}  
\begin{equation}\label{53}
\mathrm{d}s^2=-\left(1-\epsilon \right) \mathrm{d}t^2+\mathrm{d}\rho^2+\eta^2 \rho^2 \mathrm{d}\varphi^2+\mathrm{d}z^2.
\end{equation} 

Introducing a spherically coordinates transformations to the above metric and considering the equatorial plane, for the optical metric it follows
\begin{equation}
\mathrm{d}t^2=\frac{\mathrm{d}r^2}{\left(1-\epsilon \right)}+\frac{\eta^2 r^2  \mathrm{d}\varphi^2}{\left(1-\epsilon \right)}.
\end{equation}

We now introduce a new coordinate $r^{\star}$, thereby a new function $f(r^{\star})$:
\begin{equation}
\mathrm{d}r^{\star}=\frac{\mathrm{d}r}{\sqrt{1-\epsilon}}, \,\,\,f(r^{\star})=\frac{\eta r} {\sqrt{1-\epsilon}}.
\end{equation}

Morover we show that the corresponding Gaussian optical curvature vanishes also in this case i.e. $K=0$. We need to apply the GBT but first let us see that for very large $R$,  the optical metric gives
\begin{equation}
\mathrm{d}t=\frac{\eta R}{\sqrt{1-\epsilon}} \,\mathrm{d}\,\varphi
\end{equation}

Using the optical metric (49), one can show in a similar way the following relation for the geodesic curvature $ \kappa(C_{R}) \to R^{-1} \sqrt{1-\epsilon} $, and consequently $\kappa(C_{R})\mathrm{d}t=\eta \,\mathrm{d}\,\varphi$. From the GBT we find
\begin{eqnarray}\notag
\iint\limits_{D_{R}}K\,\mathrm{d}S&+&\oint\limits_{C_{R}}\kappa\,\mathrm{d}t\overset{{R\to \infty}}{=}\iint\limits_{S_{\infty}}K\,\mathrm{d}S \\
&+& \eta\, \int\limits_{0}^{\pi+\hat{\alpha}}\mathrm{d}\varphi=\pi.
\end{eqnarray}

Hence, for the deflection angle the last equation we derive the following result
\begin{equation}
\hat{\alpha} \simeq 4\mu \pi+\mathcal{O}(\mu^2).
\end{equation}

As expected, the LSBE parameter for a FTL 4-vector case is not relevant for light deflection, which is different from the FSL 4-vector case.

\subsection{Rotating cosmic strings}

In a similar way as in the FSL 4-vector case, we shall introduce a rotating cosmic string into the effective metric \eqref{53} In that case we find
\begin{equation} \label{59}
\mathrm{d}s^2=-\left(1-\epsilon \right) \left(\mathrm{d}t+a \mathrm{d}\varphi \right)^2+\mathrm{d}\rho^2+\eta^2 \rho^2 \mathrm{d}\varphi^2+\mathrm{d}z^2.
\end{equation} 

Now we will try to show that the condition $\xi_{\mu}(x) \xi^{\mu}(x) =\text{const}$ is indeed satisfied. To do so, let us first choose the following tetrads for our metric \eqref{59}
\begin{equation}
{e^a}_{\mu}(x)=\begin{pmatrix}
 \sqrt{1-\epsilon} & 0 &  0& 0\\ 
 0&  1&  0& 0\\ 
a \sqrt{1-\epsilon} &  0&  \eta \rho & 0\\ 
 0& 0 &  0& 1
\end{pmatrix}.
\end{equation}

Then by writing the four vector $\xi_{\mu}(x)={e^a}_{\mu}(x) \xi_{a}$, and choosing for generally $\xi_{a}=(\zeta,\sigma,\gamma,\delta)$, one can show that $\xi_{\mu}(x) \xi^{\mu}(x) =-\zeta^2+\sigma^2+\gamma^2+\delta^2=\text{const}$. Furthermore if we choose, $\xi_{a}=(1,0,0,0)$, we find $\xi_{\mu}(x) \xi^{\mu}(x) =-1=\text{const}$. 

Thus, by considering the spherical coordinate transformations, we find
\begin{equation}
\mathrm{d}s^2=-\beta\, (\mathrm{d}t+a \mathrm{d}\varphi)^2+\mathrm{d}r^2+r^2 \mathrm{d}\theta^2+\eta^2 r^2  \sin^2{\theta} \mathrm{d}\varphi^2
\end{equation} 
where $\beta=1-\epsilon$.  This leads to the considerably simpler Randers type metric
\begin{equation}\label{16-3}
F\left(r,\varphi,\frac{\mathrm{d}r}{\mathrm{d}t},\frac{\mathrm{d}\varphi}{\mathrm{d}t}\right)=\sqrt{\frac{1}{\beta}\left(\frac{\mathrm{d}r}{\mathrm{d}t}\right)^2+\frac{\eta^2 r^2}{\beta} \left(\frac{\mathrm{d}\varphi}{\mathrm{d}t}\right)^2}-a\frac{\mathrm{d}\varphi}{\mathrm{d}t}.
\end{equation}

The effective Gaussian curvature gives:
\begin{equation}
K=-\frac{12a}{r}f(r,\varphi,\eta,\beta)
\end{equation}
where 

\begin{widetext}
\begin{eqnarray}\notag
f(r,\varphi,\eta,\beta)&=&\frac{\sin^3 \varphi}{\left( \frac{\cos^2 \varphi}{\beta}+\frac{r^2 \eta^2 \sin^4 \varphi}{b^2 \beta}\right)^{7/2}b^7 \beta^2}\Big[-\frac{\sin^{11}\varphi \eta^6 r^5}{24}+\frac{b^2 r^3 \eta^2  \sin^9 \varphi}{8}+\frac{r^3 \eta^2  \cos^{2}\varphi b^2 (\eta^2+27) \sin^7 \varphi}{24} \\
&-&\frac{3 b^3 r^2 \eta^2  \cos^2 \varphi \sin^6 \varphi }{4}+\Big(\frac{5 b^2 r^3 \eta^2 \cos^4 \varphi}{4}-\frac{\cos^2 \varphi b^4 r}{2}\Big)\sin^5 \varphi- \frac{3 r^2 b^3 \eta^2 \cos^4 \varphi \sin^4 \varphi }{2}\\
&+& \frac{17 r (\eta^2 -\frac{33}{17}) \cos^4 \varphi b^4 \sin^3 \varphi}{24}+\frac{\cos^4 \varphi \sin^2 \varphi b^5}{2} -\frac{5 b^4 r \sin \varphi \cos^6 \varphi}{4}+b^5 \cos^6 \varphi \Big].  
\end{eqnarray}
\end{widetext}

We can now integrate with respect to the radial coordinate first and then make a Taylor series expansion around $\mu$ and $\epsilon$, for the retrograde light ray we find
\begin{equation}
\int\limits_{0}^{\pi}\int\limits_{\frac{b}{\sin \varphi}}^{\infty}\frac{12  a}{r}f(r,\varphi,\eta,\beta)\,\sqrt{\det \bar{g}}\,\mathrm{d}r\,\mathrm{d}\varphi=\frac{3 \pi a \mu}{2b}\left(1-\frac{\epsilon }{2}\right).
\end{equation}

While for the prograde light ray we find zero contribution
\begin{equation}
\int\limits_{0}^{\pi}\int\limits_{\frac{b}{\sin \varphi}}^{\infty}\frac{12  a}{r}f(r,\varphi,\eta,\beta)\,\sqrt{\det \bar{g}}\,\mathrm{d}r\,\mathrm{d}\varphi=0.
\end{equation}

For the total deflection angle, we find the following results for the retrograde case
\begin{equation}\label{68}
\hat{\alpha}_{ret}=4 \mu \pi+\frac{3 \pi a \mu}{2b}\left(1-\epsilon \right), 
\end{equation}
and for the prograde case
\begin{equation}
\hat{\alpha}_{prog}=4 \mu \pi.
\end{equation}

Hence, since the rotating cosmic string parameter $a$ is proportional to the angular momentum $J$, the last two terms in Eq. \eqref{68} can be considered as second order terms $\mu^2$. Therefore, by neglecting these terms we find 
\begin{equation}\label{70}
\hat{\alpha}=4 \mu \pi.
\end{equation}

\section{Geodesics Equations}
\subsection{Effective metric for FSL 4-vector case}
We can apply now the variational principle $\delta \int \mathcal{L} \,\mathrm{d}s=0$, to calculate the deflection angle in the stationary spacetime metric \eqref{22}. The Lagrangian can be written as \cite{Boyer}
\begin{equation}
\mathcal{L}=-\frac{1}{2}\left(\dot{t}+a \dot{\varphi}\right)^2+\frac{\dot{r}^2}{2}+\frac{1}{2} r(s)^2\left(\dot{\theta}^2+\eta^2 \alpha \sin^{2}\theta \dot{\varphi}^2\right)
\end{equation}

We can simplify further this problem by choosing $\theta =\pi/2$. Next, we introduce two constants of motion, say $l$, and $\gamma$, given by
\begin{eqnarray}
p_{\varphi}&=&\frac{\partial \mathcal{L}}{\partial \dot{\varphi}}=-\left(\dot{t}+a \dot{\varphi}\right)a+\eta^2 \alpha r(s)^2 \dot{\varphi} =l   \\
p_{t}&=&\frac{\partial \mathcal{L}}{\partial \dot{t}}=-\left(a \dot{\varphi}+\dot{t}\right)=-\gamma.
\end{eqnarray}

We change the coordinates by using $r=1/u(\varphi)$, which leads to the following identiy 
\begin{equation}
\frac{\dot{r}}{\dot{\varphi}}=\frac{\mathrm{d}r}{\mathrm{d}\varphi}=-\frac{1}{u^2}\frac{\mathrm{d}u}{\mathrm{d}\varphi}.
\end{equation}

Morover we make clear that the angle $\varphi$ is measured from the point of closest approach i.e. $u=u_{max}=1/r_{min}=1/b$ \cite{lorio}. Hence, we can choose for the first and second constant  $\gamma=1$ and $l=\eta \sqrt{\alpha} \,b$, respectively. This leads to the following equation
\begin{widetext}
\begin{eqnarray}\label{75}
\frac{1}{2 u^4}\left( \frac{\mathrm{d}u}{\mathrm{d}\varphi}\right)^2+\frac{\eta^2 \alpha }{2 u^2}-\frac{1}{2}\frac{\left(\eta^2 \alpha-a^2 u^2-\eta \sqrt{\alpha}b u^2 a \right)^2}{u^4 \left(a+\eta \sqrt{\alpha}b\right)^2}-\frac{a \left(\eta^2 \alpha-a^2 u^2-\eta \sqrt{\alpha}b u^2 a \right)}{u^2 \left(a+\eta \sqrt{\alpha}b\right)}-\frac{a^2}{2}=0
\end{eqnarray}
\end{widetext}

We can solve the above differential equation \eqref{75} using a perturbation method.  The solution of this differential equation in leading order terms can be written in the form
\begin{equation}
\Delta \varphi =\pi+\hat{\alpha},
\end{equation}
where $\hat{\alpha}$ is the deflection angle. One can show that the above result can be written as \cite{weinberg}
\begin{equation}
\hat{\alpha}=2|\varphi(u_{max})-\varphi_{\infty}|-\pi.
\end{equation}

Solving for $\mathrm{d}u/\mathrm{d}\varphi$ and using a Taylor series expansion around $\mu$, $a$, and $\epsilon$, leads to the following integral 
\begin{equation}
\varphi= \int_0 ^{1/b}  A(u,\mu,a,\epsilon,b)  \mathrm{d}u.
\end{equation}

The function $A(u,\mu,a,\epsilon,b)$ is given by
\begin{equation}
A(u,\mu,a,\epsilon,b)=\frac{b\left[2a (\epsilon -1)(8 \mu+1)+\zeta \right] }{2\sqrt{(1-b^2 u^2)b^2}\left(b^2 u^2-1\right)},
\end{equation}
where 
\begin{eqnarray}\notag
\zeta &=&\left((4\mu+1)\epsilon -8 \mu-2\right)  -(\epsilon -2)(4 \mu+1)u^2 b^3.
\end{eqnarray}

We find the following result for the deflection angle in leading order terms
\begin{equation}
\hat{\alpha} \simeq 4 \mu \pi -\frac{\epsilon \pi}{2}-2 \pi \epsilon \mu.
\end{equation}

Thus, we have recovered the deflection angle which corresponds to to the static case given by Eq. \eqref{51}.

\subsection{Effective metric for FTL 4-vector case}

Using similar approach we can calculate the deflection angle in the stationary spacetime metric (54). The Lagrangian can be written as
\begin{equation}
\mathcal{L}=-\frac{\beta}{2}\left(\dot{t}+a \dot{\varphi}\right)^2+\frac{\dot{r}^2}{2}+\frac{1}{2} r(s)^2\left(\dot{\theta}^2+\eta^2  \sin^{2}\theta \dot{\varphi}^2\right)
\end{equation}

Then in the equatorial plane we find
\begin{eqnarray}
p_{\varphi}&=&\frac{\partial \mathcal{L}}{\partial \dot{\varphi}}=-\beta \left(\dot{t}+a \dot{\varphi}\right)a+\eta^2  r(s)^2 \dot{\varphi} =l   \\
p_{t}&=&\frac{\partial \mathcal{L}}{\partial \dot{t}}=-\beta \left(a \dot{\varphi}+\dot{t}\right)=-\gamma.
\end{eqnarray}

Without loss of generality, we can choose $\gamma=1$ and $l=(\eta b)/ \sqrt{\beta}$. The following differential equation can be obtained 
\begin{widetext}
\begin{equation}\label{83}
\frac{1}{2 u^4}\left( \frac{\mathrm{d}u}{\mathrm{d}\varphi}\right)^2+\frac{\eta^2 }{2 u^2}-\frac{1}{2}\frac{\left(\eta^2-a^2 \beta u^2-\eta \sqrt{\beta}\,b u^2 a \right)^2}{u^4 \beta  \left(a+(\eta\,b)/ \sqrt{\beta}\right)^2}-\frac{a\left(\eta^2-a^2 \beta u^2-\eta \sqrt{\beta}\,b u^2 a \right)}{\left(a+(\eta\,b)/ \sqrt{\beta}\right)}-\frac{a^2\beta }{2}=0.
\end{equation}
\end{widetext}

This result leads to the following integral
\begin{equation}
\varphi=\int_0 ^{1/b}  B(u,\mu,a,\epsilon,b)  \mathrm{d}u,
\end{equation}
where the function $B(u,\mu,a,\epsilon,b)$ is given by
\begin{equation}
B(u,\mu,a,\epsilon,b)=\frac{b\left[a (\epsilon -2)(8 \mu+1)+\Xi \right] }{2\sqrt{(1-b^2 u^2)b^2}\left(b^2 u^2-1\right)},
\end{equation}
where 
\begin{eqnarray}\notag
\Xi &=& b \left(2 u^2 b^2(4\mu+1) -2b (4\mu+1)\right)
\end{eqnarray}

Working in a similar fashion as in the FSL 4-vector case, the solution of equation \eqref{83} is written as
\begin{equation}
\Delta \varphi =\pi+\hat{\alpha},
\end{equation}
with the deflection angle $\hat{\alpha}$ given by
\begin{equation}
\hat{\alpha} \simeq 4 \mu \pi.
\end{equation}

This result is consistent with the Eq. \eqref{70} found by the GB method.

\section{Conclusion}

In this paper, we have first computed the deflection angle by virtue of a cosmic string having the LSBE. To this end, we have applied the GBT to the effective-optical metric of the static cosmic string spacetime for a FSL 4-vector case. The first term of the deflection angle (18) was found to be the ordinary deflection angle by a static cosmic string. However, it has been shown that Lorentz symmetry breaking, which is parametrized by $\epsilon$ decreases the deflection angle. On the other hand, for a FTL 4-vector case with the LSBE, the deflection angle remains unchanged, remarkably. Next, we have extended our results to rotating cosmic strings. We have derived their corresponding both Randers type effective-cosmic string optical metrics and the effective Gaussian optical curvatures. We then have constructed the osculating Riemannian manifolds and applied the GBT to those obtained optical metrics. We have deduced from our results that the deflection angle is not affected by the rotation of the cosmic string in leading order terms. Thus, the results obtained by the GB method are in perfect agreement with the geodesics computations in the leading order terms.

It is worth noting that the main outcome of this study is that LSBE plays an important role on the light deflection in a cosmic string spacetime. From this point of view, the latter remark is the newest contribution to the subject of gravitational lensing. Besides, LSBE can be considered in the future observations about gravitational lensing. Moreover, GB method is a powerful theoretical technique for finding an exact result of the deflection angle. Because, it evaluates the associated deflection angle integral in the domain that connotates an infinite domain bounded by the light ray. In conclusion, GB method reveals that light deflection can be seen as a \textit{partially topological effect} of the spacetime geometry, which is very recently discussed in [Physics Letters A 381 (2017) 1764-1772] \cite{Isabel}.

\begin{acknowledgments}
We thank Marcus C. Werner for his helpful comments and constructive suggestions. This work was supported by the Chilean FONDECYT Grant No. 3170035 (A\"{O}).
\end{acknowledgments}

\end{document}